\documentstyle[preprint,eqsecnum,aps]{revtex}
\tightenlines
\newcommand{\be}{\begin{equation}}
\newcommand{\ee}{\end{equation}}
\newcommand{\ba}{\begin{eqnarray}}
\newcommand{\ea}{\end{eqnarray}}

\begin{document}
\draft

\title{Entropies of Rotating Charged Black Holes \\ from
Conformal Field Theory at Killing Horizons}
 \author{Jiliang Jing $^{*\ a\ b }$\footnotetext[1]
 {email: jljing@hunnu.edu.cn} \ \
 \ \ Mu-Lin Yan $^{\dag\ b}$\footnotetext[2]
 {email: mlyan@ustc.edu.cn}}
\address{a) Physics Department and Institute of Physics , Hunan Normal
University,\\ Changsha, Hunan 410081, P. R. China;  \\ b)
Department of Astronomy and Applied Physics, University of Science
and Technology of China, \\ Hefei, Anhui 230026, P. R. China}

\maketitle
\begin{abstract}

The covariant phase technique is used to compute the constraint
algebra of the stationary axisymmetric charged  black hole. A
standard Virasoro subalgebra with corresponding central charge is
constructed  at a Killing horizon with Carlip's boundary
conditions. For the Kerr-Newman black hole and the Kerr-Newman-AdS
black hole, the density of states determined by conformal fields
theory methods yields the statistical entropy which agrees with
the Bekenstein-Hawking entropy.
 \vspace*{1.5cm}
\end{abstract}


\section{INTRODUCTION}
\label{sec:intro} \vspace*{0.5cm} The statistical mechanical
description of the Bekenstein-Hawking black hole entropy
\cite{Bekenstein72}- \cite{Hawking74} in terms of microscope
states is an outstanding open question and much effort has been
concentrated on the problem for some years
\cite{Hooft85}-\cite{Mann96}. Success seems to come with the paper
of Strominger and Vafa \cite{Strominger96} which was followed by a
host of others. It is well known since the work of Brown and
Henneaux \cite{Brown86} that a asymptotic symmetry group of
AdS$_3$ is generated by a Virasoro algebra, and that therefore any
consistent quantum theory of gravity on AdS$_3$ is conformal field
theory. Using the result Strominger \cite{Strominger98} calculated
the entropy of black holes whose near-horizon geometry is locally
$AdS_3$ from the asymptotic growth of states. Precise numerical
agreement with the Bekenstein-Hawking area formula for the entropy
was found. In light of the work, one could statistically
reinterpret the black hole entropy by establishing a relation
conformal field theory on the boundary of related anti-de Sitter
space. In order to overcome the limitations of Strominger's
method, such as the approach can only be used for 2+1 dimensional
spacetime and it is based on an algebra of transformations at
infinity, Carlip \cite{Carlip99l} generalized
Brown-Henneaux-Strominger's approach by looking at the symmetries
of the event horizon of an (n+1)-dimensional Schwarzschild-like
black hole. This construction is valid for black hole in any
dimension. In Ref. \cite{Carlip99} Carlip re-derived the central
extension of the constraint algebra of general relativity by using
manifestly covariant phase space methods
\cite{Lee90}-\cite{Iyer95} and a boundary which is a surface that
look like a (local) Killing horizon. A  natural set of boundary
conditions leads to a Virasoro subalgebra with a calculable
central charge. Then, by means of conformal field theory method,
Carlip \cite{Carlip99} studied the statistical entropies of the
Rindler space, static de Sitter space, Taub-NUT and Taub-Bolt
spaces, and 2-dimensional dilaton gravity. However, at the moment
the question whether or not the covariant phase space approach can
be used for the stationary axisymmetric charged black holes which
are described by solutions of the Einstein-Maxwell equations, such
as the Kerr-Newman black hole and the Kerr-Newman-AdS black hole,
still remains open. The aim of this paper is to settle the
question.

The  paper is organized  as follows: In Sec. II, by using the
covariant phase techniques we extend Carlip's investigation
\cite{Carlip99} for vacuum case ${\mathbf{L}}_{a_1a_2\cdots
a_{n}}= \frac{1}{16\pi G}\epsilon_{a_1a_2\cdots a_{n}}R $ to a
case including a cosmological term and electromagnetic fields,
i.e., the Lagrangian n-form is described by
${\mathbf{L}}_{a_1a_2\cdots a_{n}}= \frac{1}{16\pi
}\epsilon_{a_1a_2\cdots
a_{n}}\left[\frac{1}{G}(R-2\Lambda)+F^{ab}F_{ab}\right].$ In Sec.
III, the standard Virasoro subalgebras with corresponding central
charges are constructed for the Kerr-Newman black hole and the
Kerr-Newman-AdS black hole. The statistical entropies for these
objects are then calculated by using Cardy formula. Some
discussions and summaries are presented in the last section.

\section{Algebra of diffeomorphism} \vspace*{0.5cm}

Lee, Wald, and Iyer \cite{Lee90} \cite{Wald93} \cite{Iyer94}
\cite{Iyer95} showed that the variation of the Lagrangian defines
the equation of motion n-form ${\mathbf{E}}$ and the symplectic
potential (n-1)-form ${\mathbf{\Theta}}$ via the equation
$
\delta  {\mathbf{L}}={\mathbf{E}} \delta\phi+d {\mathbf{\Theta}} $
,  where ${\mathbf{L}}$ is an n-form,  $ {\mathbf{E}}
\delta\phi={\mathbf{E}}^{ab}_{g}\delta g_{ab}+{\mathbf{E}} _{\psi}
\delta \psi, $    $\phi =(g_{ab}, \psi)$ denotes an arbitrary
collection of dynamical fields, and the equations of motion are
taken to be ${\mathbf{E}}^{ab}_{g}=0$ and ${\mathbf{E}}_{\psi}=0.
$ Let $\xi ^a$ be any smooth vector fields on the spacetime
manifold ${\mathbf{M}}$, i. e., $\xi ^a$ is the infinitesimal
generator of a diffeomorphism, we  can define a Noether current
(n-1)-form  as \cite{Iyer94}\cite{Iyer95}
\begin{equation}
{\mathbf{J}}[\xi]={\mathbf{\Theta}} [\phi, {\mathcal{L}}_{\xi}
\phi]-{\mathbf{\xi}} \cdot {\mathbf{L}}, \label{j1}
\end{equation}
here and hereafter the ``central dot" denotes the contraction of
the vector field $\xi ^a$ into the first index of the differential
form. By using the equations of motion a standard calculation
\cite{Lee90} shows that ${\mathbf{J}}$ is closed for all $\xi ^a$,
i.e., $d {\mathbf{J}}=0$. Then we have \cite{Iyer94}
\begin{equation}
{\mathbf{J}}=d {\mathbf{Q}},\label{dQ}
\end{equation}
where ${\mathbf{Q}}$ is a Noether charge (n-2)-form. From the
variation of Noether current (n-1)-form, we know that the
symplectic current  (n-1)-form $\omega [\phi, \delta
_1\phi, \delta _2\phi]=\delta _2{\mathbf{\Theta}} [\phi, \delta
_1\phi]-\delta _1{\mathbf{\Theta}} [\phi, \delta _2\phi] $
 can be expressed as \cite{Lee90}
\begin{equation}
\omega [\phi, \delta \phi, {\mathcal{L}}_{\xi}\phi]=\delta
{\mathbf{J}}[\xi]-d({\mathbf{\xi}} \cdot {\mathbf{\Theta}} [\phi,
\delta \phi]),
\end{equation}
and Hamilton's equation of motion is given by
\begin{equation}
\delta H[\xi]=\int _C \omega [\phi, \delta \phi,
{\mathcal{L}}_{\xi}\phi]=\int _C[\delta
{\mathbf{J}}[\xi]-d({\mathbf{\xi}} \cdot {\mathbf{\Theta}} [\phi,
\delta \phi])]. \label{dh}
\end{equation}
By using Eq. (\ref{dQ}) and Carlip's boundary conditions listed in
Appendix A and defining a (n-1)-form ${\mathbf{B}}$ as
\begin{equation}
  \delta \int _{\partial C} {\mathbf{\xi}} \cdot
 {\mathbf{ B}}[\phi]=\int _{\partial C}{\mathbf{\xi}} \cdot
 {\mathbf{\Theta}} [\phi. \delta \phi],\label{bq}
\end{equation}
the Hamiltonian can be expressed as \cite{Carlip99}
\begin{equation}\label{H}
H[\xi]=\int _{\partial C}({\mathbf{Q}}[\xi]-{\mathbf{\xi}}\cdot
{\mathbf{B}}[\phi]).
\end{equation}
It is well-known that the Poisson bracket forms a standard
``surface deformation algebra"  \cite{Brown86} \cite{Carlip99}
\begin{equation}\label{algeb}
  \{H[\xi_1], H[\xi _2]\}=H[\{\xi_1, \xi_2\}]+K[\xi_1, \xi_2],
\end{equation}
where the central term $K[\xi_1, \xi_2]$ depends on the dynamical
fields only through their boundary values.

In this paper, we focus our attention to stationary axisymmetric
charged black holes. So we take the Lagrangian n-form as
\begin{equation}\label{L2}
{\mathbf{L}}_{a_1 a_2 \cdots a_n}=\frac{1}{16 \pi} \epsilon_{a_1
a_2 \cdots a_n}\left[\frac{1}{G}(R-2 \Lambda)+F^{a b}F_{a
b}\right],
\end{equation}
where $\epsilon_{a_1 a_2 \cdots a_n}$ is a volume element (a
continuous non-vanishing n-form), $\Lambda$ is the cosmological
constant, and $F_{ab}$ is the electromagnetic field strength
tensor. By using the infinitesimal generator of a diffeomorphism,
${\mathbf{\xi}}$, we know that the symplectic potential (n-1)-form
is given by
\begin{equation}\label{syp4}
{\mathbf{\Theta}}_{a_1 a_2 \cdots a_{n-1}}[g,
{\mathcal{L}}_{\xi}g]=\frac{1}{4 \pi }\epsilon_{c a_1 a_2 \cdots
a_{n-1}}\left\{\frac{1}{2G}(\nabla _e \nabla ^{[e}\xi ^{c]}+R_e^c
\xi^e) +F^{d c}\left[F_{e d}\xi^e+(\xi^e A_e)_{;d}\right]\right\}.
\end{equation}
Eqs. (\ref{j1}) and (\ref{syp4}) yields
\begin{eqnarray}\label{j2}
{\mathbf{J}}_{a_1 a_2 \cdots a_{n-1}}&=&\frac{1}{8 \pi
G}\epsilon_{c a_1 a_2 \cdots a_{n-1}}[\nabla _e \nabla ^{[e}\xi
^{c]}+(R_e^c-\frac{1}{2}\delta^c_eR+\delta^c_e \Lambda)
\xi^e]\nonumber \\ & & -\frac{1}{4\pi}\epsilon
_{ca_1a_2...a_{n-1}}\left[\frac{1}{4}F^{b d}F_{b d}\delta
^c_e-F^{c d}F_{e d}\right]\xi^e+\frac{1}{4\pi}
\epsilon_{ca_1a_2...a_{n-1}}F^{e c}(\xi^dA_d)_{;e} \nonumber
\\&=&\frac{1}{4 \pi}\epsilon_{c a_1 a_2
\cdots a_{n-1}}\left[\frac{1}{2 G}\nabla _e \nabla ^{[e}\xi
^{c]}+\nabla_e (\nabla^{[e}A^{c]}A_d\xi^d)\right],\label{j3}
\end{eqnarray}
in above calculation, we used the Einstein-Maxwell field equations
in which the energy-momentum tensors is given by
$\frac{1}{4\pi}\left[\frac{1}{4}F^{b c}F_{b c}\delta ^d_e-F^{d c
}F_{e c}\right]$.

From Eqs. (\ref{dQ}) and (\ref{j3}) we have
\begin{equation}\label{Q1}
{\mathbf{Q}}_{a_1 a_2\cdots
a_{n-2}}=-\frac{1}{4\pi}\epsilon_{bca_1...a_{n-2}}\left[\frac{1}{4
G}\nabla^{b}\xi^{c}+(\nabla^b A^c)A_e\xi^e\right].
\end{equation}

For a  stationary axisymmetric charged black hole (such as the
Kerr-Newman black hole and the Kerr-Newman AdS/dS black hole), the
electromagnetic potential $ A_a $, the electromagnetic field
tensors $F^{03}$, and the Killing vector can be expressed
respectively as
\begin{eqnarray}
A_a&=&(A_0, \ 0, \ 0, \ A_3) \nonumber \\ F^{0 3}&=&-F^{3 0}=0.
\label{Af}\\
 \chi^a_H&=&\chi^{(t)}_H+\chi^{(\varphi)}_H=(1, \ 0,
\ 0, \ \Omega_H), \label{kl}
\end{eqnarray}
where the vector $\chi^{(t)}_H$ correspond to time translation
invariance, $\chi^{(\varphi)}_H$ to rotational symmetry, and
$\Omega_H=-(g_{t\varphi}/g_{\varphi\varphi})_H$ is the angular
velocity of the black hole.

Using Eqs. (\ref{Af}), (\ref{kl}), (\ref{xi}), and (\ref{vol}) it
is easy to show that
\begin{eqnarray} & &\frac{1}{4\pi}\epsilon_{bca_1...a_{n-2}}
(\nabla^b A^c)A_e\xi^e
 \rightarrow  0. \ \ \ \ \ \ \ \ \ \ \ \  \hbox{ at the horizon}
\end{eqnarray}
Then,  Eq. (\ref{Q1}) is reduced to
\begin{equation}\label{QQ}
Q_{a_1a_2...a_{n-2}}=-\frac{1}{16\pi
G}\epsilon_{bca_1a_2...a_{n-2}}\nabla^b\xi^c.
\end{equation}

Denoting by $\delta _{\xi}$ the variation corresponding to
diffeomorphism generated by $\xi$, for the Noether current
${\mathbf{J}}[\xi]$ we have
\begin{equation}\label{dj}
\delta_{\xi_2}{\mathbf{J}}[\xi_1]=\xi_2 d {\mathbf{J}}[\xi_1]+d
(\xi_2 \cdot {\mathbf{J}}[\xi_1] )= d[\xi_2
({\mathbf{\Theta}}[\phi, {\mathcal{L}}_{\xi_{1}}
\phi]-\xi_{1}\cdot {\mathbf{L}})].
\end{equation}
Substituting Eq. (\ref{dj}) into Eq. (\ref{dh}) and using Eq.
(\ref{syp4}) we get
\begin{eqnarray}\label{dh1}
\delta _{\xi_2}H[\xi_1]&=&\int_C\left(\delta
_{\xi_2}{\mathbf{J}}[\xi_1]-d(\xi_1{\mathbf{\Theta}}[\phi,
\delta_{\xi_2}\phi ])\right) \nonumber \\ &=&\int_{\partial
C}\left(\xi_2{\mathbf{\Theta}}[\phi, {\mathcal{L}}_{\xi_1}\phi
]-\xi_1{\mathbf{\Theta}}[\phi, {\mathcal{L}}_{\xi_2}\phi
]-\xi_2\xi_1{\mathbf{L}}\right) \nonumber \\&=&\frac{1}{16\pi G}
\int _{\partial C}\epsilon_{bca_1...a_{n-2}}\left[
\xi^b_2\nabla_d(\nabla^d\xi^c_1-\nabla^c\xi^d_1)-
\xi^b_1\nabla_d(\nabla^d\xi^c_2-\nabla^c\xi^d_2)\right]\nonumber
\\
& &+  \frac{1}{8\pi}\int _{\partial C}\epsilon_{bca_1...a_{n-2}}
\left\{\xi_2^bF^{d c}\left[F_{e d}\xi_1^e+(\xi_1^e
A_e)_{;d}\right]-\xi_1^bF^{d c}\left[F_{e d}\xi_2^e+(\xi_2^e
A_e)_{;d}\right] \right\}\nonumber
\\ & &-
\frac{1}{16\pi G}\int _{\partial C}\epsilon_{bca_1...a_{n-2}}
\left[2R^c_d(\xi^b_1\xi^d_2-\xi^b_2\xi^d_1)+\xi^b_2\xi^c_1
{\mathbf{L}}\right].
\end{eqnarray}
At the horizon, by using Eqs.  (\ref{Af}), (\ref{kl}) and
(\ref{cond})- (\ref{vol}) we know
\begin{eqnarray} & &\int _{\partial C}\epsilon_{bca_1...a_{n-2}}
\xi^b_2\xi^c_1 {\mathbf{L}} \nonumber \\&=&\int _{\partial
C}\hat{\epsilon} _{a_1...a_{n-2}}{\mathbf{L}}\left[\frac{|\chi|}
{\rho}{\mathcal{T}}_2\rho_c+
\left(\frac{\rho}{|\chi|}+t\cdot\rho\right){\mathcal{R}}_2
\chi_c\right]
({\mathcal{T}}_1\chi^c+{\mathcal{R}}_1\rho^c)\nonumber \\ &=&\int
_{\partial C}\hat{\epsilon} _{a_1...a_{n-2}}{\mathbf{L}} \left[
\frac{|\chi|}{\rho} {\mathcal{T}}_2{\mathcal{R}}_1\rho^2+
\left(\frac{\rho}{|\chi|}+
t\cdot\rho\right){\mathcal{R}}_2{\mathcal{T}}_1\chi^2\right]
\nonumber
\\&=&0,
\end{eqnarray}
\begin{eqnarray}
& &\int _{\partial C}\epsilon_{bca_1...a_{n-2}}
2R^c_d(\xi^b_1\xi^d_2-\xi^b_2\xi^d_1)\nonumber \\ &=&\int
_{\partial C}\hat{\epsilon}
_{a_1...a_{n-2}}R^c_d\left(\frac{1}{\kappa}\frac{\chi^2}{\rho^2}
\right)\left[\frac{|\chi|}{\rho}\rho_c\rho^d-\left(
\frac{\rho}{|\chi|}+t\cdot\rho\right)\chi_c\chi^d\right]
({\mathcal{T}}_1D{\mathcal{T}}_2-{\mathcal{T}}_2
D{\mathcal{T}}_1)\nonumber
\\ &=&0,
\end{eqnarray}
and
\begin{eqnarray} &
&\xi^b\epsilon _{bca_1a_2...a_{n-2}}F^{d c}\left[F_{e
d}\xi^e+(\xi^e A_e)_{;d}\right]\nonumber \\ &=&\xi^b\epsilon
_{bca_1a_2...a_{n-2}}F^{d c}\delta_{\xi}A_d\nonumber \\
&=&\hat{\epsilon}
_{a_1a_2...a_{n-2}}\left[\frac{|\chi|}{\rho}{\mathcal{T}}\rho_c+
\left(\frac{\rho}{|\chi|}+t\cdot\rho\right)
{\mathcal{R}}\chi_c\right]F^{d c}\delta_{\xi}A_d\nonumber \\ &=&0.
\label{second}
\end{eqnarray}
Therefore, Eq. (\ref{dh1}) can be rewritten as
\begin{eqnarray}\label{dh1a}
\delta _{\xi_2}H[\xi_1]&=&\frac{1}{16\pi G} \int _{\partial
C}\epsilon_{bca_1...a_{n-2}}\left[
\xi^b_2\nabla_d(\nabla^d\xi^c_1-\nabla^c\xi^d_1)-
\xi^b_1\nabla_d(\nabla^d\xi^c_2-\nabla^c\xi^d_2)\right].
\end{eqnarray}
Since the ``bulk" part of the generator $H[\xi_1]$ on the left
side vanishes on shell, we can interpret the left side of Eq.
(\ref{dh1}) the variation of the boundary term $J$, i.e., $\delta
_{\xi_2}J[\xi_1]$. On the other hand, the change in $J[\xi_1]$
under a surface deformation generated by $J[\xi_2]$ can be
precisely described by Dirac bracket $\{J[\xi_1], j[\xi_2]\}^*$
\cite{Carlip99}, that is,
\begin{equation}\label{dj3}
\delta_{\xi_2}J[\xi_1]=\{J[\xi_1], J[\xi_2]\}^*=\frac{1}{16\pi G}
\int _{\partial C}\epsilon_{bca_1...a_{n-2}}\left[
\xi^b_2\nabla_d(\nabla^d\xi^c_1-\nabla^c\xi^d_1)-
\xi^b_1\nabla_d(\nabla^d\xi^c_2-\nabla^c\xi^d_2)\right].
\end{equation}
Above discussions and Eq. (\ref{algeb}) show the following
relation on shell
\begin{equation}\label{algeb1}
\{J[\xi_1], J[\xi_2]\}^*=J[\{\xi_1, \xi_2\}]+K[\xi_1, \xi_2],
\end{equation}

Substituting Eqs. (\ref{xi}), (\ref{rt}), and (\ref{vol}) into Eq.
(\ref{dj3}) we find that
\begin{eqnarray}\label{dj5}
\{J[\xi_1], J[\xi_2]\}^*&=& -\frac{1}{16\pi G}\int_{\partial
C}\hat{\epsilon}_{a_1 \cdots
a_{n-2}}\left[\frac{1}{\kappa}({\mathcal{T}}_1D^3{\mathcal{T}}_2
-{\mathcal{T}}_2D^3{\mathcal{T}}_1)-2\kappa
({\mathcal{T}}_1D{\mathcal{T}}_2
-{\mathcal{T}}_2D{\mathcal{T}}_1)\right].
\end{eqnarray}
It is also easy to show that
\begin{equation}\label{xixi}
\{\xi_1,\xi_2\}^a=({\mathcal{T}}_1D{\mathcal{T}}_2
-{\mathcal{T}}_2D{\mathcal{T}}_1)\chi^a+\frac{1}
{\kappa}\frac{\chi^2}{\rho^2} D({\mathcal{T}}_1D{\mathcal{T}}_2
-{\mathcal{T}}_2D{\mathcal{T}}_1)\rho^a.
\end{equation}
On the other hand, the integrand of the right hand of  Eq.
(\ref{bq}) can be expressed as
\begin{equation}\label{syp4a}
\xi^b{\mathbf{\Theta}}_{ba_1...a_{n-2}}=\frac{1}{4 \pi
}\xi^b\epsilon_{bc a_1  \cdots a_{n-2}}\left\{\frac{1}{2G}(\nabla
_e \nabla ^{[e}\xi ^{c]}+R_e^c \xi^e) +F^{d c}\left[F_{e
d}\xi^e+(\xi^e A_e)_{;d}\right]\right\}.
\end{equation}
The first two terms in the right hand of Eq. (\ref{syp4a}) can be
treated as Carlip did in Ref. \cite{Carlip99}. And by using Eqs.
(\ref{second}) we know that the last two terms in Eq.
(\ref{syp4a}) gives no contribution to $K[\xi_1, \xi_2]$. Making
use of  Eqs. (\ref{dQ}), (\ref{QQ}), (\ref{xi}), (\ref{rt}), and
(\ref{vol}) and replacing $\xi^a$ in $J$ by $\{\xi_1, \xi_2\}^a$,
we have
\begin{equation}\label{j4}
J[\{\xi_1, \xi_2\}]=\frac{1}{16 \pi G}\int_{\partial
C}\hat{\epsilon}_{a_1 a_2\cdots
a_{n-2}}\left[2\kappa({\mathcal{T}}_1D{\mathcal{T}}_2
-{\mathcal{T}}_2D{\mathcal{T}}_1)-\frac{1}{\kappa}
D({\mathcal{T}}_1D^2{\mathcal{T}}_2
-{\mathcal{T}}_2D^2{\mathcal{T}}_1)\right].
\end{equation}
The central term can then be obtained from Eqs.(\ref{algeb1}),
(\ref{dj5}), and (\ref{j4}), which is explicitly given by
\begin{equation}\label{kk}
K[\xi_1, \xi_2]=\frac{1}{16 \pi G}\int_{\partial
C}\hat{\epsilon}_{a_1 a_2\cdots
a_{n-2}}\frac{1}{\kappa}({D\mathcal{T}}_1D^2{\mathcal{T}}_2-
D{\mathcal{T}}_2D^2{\mathcal{T}}_1).
\end{equation}

\vspace*{0.5cm}
\section{Entropy of some rotating charged black holes}

In this section, lets us study statistical-mechanical entropies of
the stationary axisymmetric black holes by using the constraint
algebra constructed in the preceding section and conformal field
theory methods.

\subsection{Entropy of the Kerr-Newman black hole}

In Boyer-Lindquist coordinates, the metric of the Kerr-Newman
black hole  takes the form \cite{Kerr63}\cite{Newman65}
\be
 ds^2 =-\frac{\Delta}{\varrho^2}
 \left[dt - a \sin^2\theta d\varphi\right]^2
 + \frac{\varrho^2}{\Delta}dr^2 + \varrho^2d\theta^2
 + \frac{\sin^2\theta }{\varrho^2}\left[a dt -
 (r^2+a^2)d\varphi \right]^2,\label{metrick}
\ee with \ba
 \varrho^2 & = & r^2 + a^2\cos^2\theta, \nonumber \\
 \Delta & = & (r-r_+)(r-r_-),
 \ea
where $r_+=r_H=M+\sqrt{M^2-Q^2-a^2}$, $r_-=M-\sqrt{M^2-Q^2-a^2}$,
the parameter $a$ is related to the angular momentum, and $M$ and
$Q$ represent the mass and electric charge of the black hole,
respectively. The metric (\ref{metrick}) is a  solution of  the
Einstein-Maxwell field equations with an electromagnetic vector
potential \ba
 {\mathbf{A}}&=&-\frac{Q r}{\varrho^2}(d t-a sin^2\theta d
 \varphi), \label{aak}
\ea and associated field strength tensor \ba
 {\mathbf{F}}=&-&\frac{Q}{\varrho^4} (r^2-a^2\cos^2\theta)
 e^0\wedge e^1\nonumber\\
 &+&\frac{Q}{\varrho^4} (r^2-a^2\cos^2\theta)e^2\wedge e^3.
\ea
 The Killing vector can be expressed as
\be
 \chi_H^a = (1, \ 0, \ 0,\  \Omega_H),
\ee where $\Omega_H=-\left(\frac{g_{t\varphi}}{g_{\varphi \varphi
}}\right)_H=\frac{ a }{r^2+a^2}$
 is the angular velocity of the black hole.
A one-parameter group of diffeomorphism satisfying Eqs.
(\ref{cond2}) and (\ref{xixi}) can be taken as
\begin{equation}\label{btk}
{\mathcal{T}}_n=\frac{1}{\kappa}exp\left[in(\kappa t + C_\alpha
(\varphi-\Omega_H t) )\right],
\end{equation}
where $C_\alpha$ is a arbitrary constant.  Substituting Eq.
(\ref{btk})  into central term (\ref{kk}) and using condition
(\ref{cond2}) we obtain
\begin{equation}\label{bkkk}
  K[{\mathcal{T}}_m, {\mathcal{T}}_n]=-\frac{iA_H}{8\pi
  G}m^3\delta_{m+n, 0},
\end{equation}
where $A_H=\int _{\partial
C}{\widehat{\epsilon}_{a_1a_2...a_{n-2}}}=4\pi(r_+^2+a^2)$ is the
area of the event horizon. Thus, Eq. (\ref{algeb1}) takes standard
form of a Virasoro algebra
\begin{equation}\label{algeb2k}
i\{J[{\mathcal{T}}_m],
J[{\mathcal{T}}_n]\}=(m-n)J[{\mathcal{T}}_{m+n}]+
\frac{c}{12}m^3\delta_{m+n, 0},
\end{equation}
with central charge $\frac{c}{12}=\frac{A_H}{8\pi G}.$ The
boundary term $J[{\mathcal{T}}_0]$ can easily be  obtained by
using Eqs (\ref{dQ}), (\ref{Q1}), and (\ref{btk}), which is given
by $ J[{\mathcal{T}}_0]=\frac{A_H}{8\pi G}. $ The number of states
with a given eigenvalue $\triangle$ of $J[{\mathcal{T}}_0]$ grows
asymptotically for large $\triangle$ as
\begin{equation}\label{brhok}
\rho(\triangle)\sim exp\left\{2\pi
\sqrt{\frac{c}{6}\left(\triangle-\frac{c}{24}\right)}
\right\}=exp\left[\frac{A_H}{4G}\right],
\end{equation}
and the statistical entropy of the Kerr-Newman black hole is
\begin{equation}\label{bsk}
\mathrm{log}\rho (\triangle)\sim \frac{A_H}{4 G},
\end{equation}
which coincides with the standard Bekenstein-Hawking entropy.

\subsection{Entropy of the Kerr-Newman-AdS black hole}

Carter \cite{Carter68} constructed the Kerr-Newman-AdS black hole
in four dimensions many years ago,  which can be explicitly given
by
\be
 ds^2 =-\frac{\Delta_r}{\varrho^2}
 \left[dt - \frac{a}{\Xi} \sin^2\theta d\varphi\right]^2
 + \frac{\varrho^2}{\Delta_r}dr^2 + \frac{\varrho^2}
 {\Delta_\theta}d\theta^2
 + \frac{\sin^2\theta \Delta_\theta}{\varrho^2}\left[a dt -
 \frac{(r^2+a^2)}{\Xi} d\varphi \right]^2, \label{metric}
\ee with \ba
 \varrho^2 & = & r^2 + a^2\cos^2\theta, \nonumber \\
 \Delta_r & = & (r^2 + a^2)(1+ l^2 r^2) - 2Mr+q^2+p^2, \nonumber \\
 \Delta_\theta & = & 1 - l^2 a^2 \cos^2\theta, \nonumber \\
 \Xi & = & 1 - l^2 a^2, \nonumber
\ea
 where the parameter $M$ is related to the mass, $a$ to the
angular momentum, $q$ is proportional to the electric charge, $p$
is proportional to the magnetic charge, and $l^2 = -\Lambda/3$ (
where $\Lambda$ is the (negative) cosmological constant).   The
event horizon is located at $r=r_+$, the largest root of the
polynomial $\Delta_r$. The metric (\ref{metric}) is a  solution of
the Einstein-Maxwell field equations with an electromagnetic
vector potential is given by \ba
 {\mathbf{A}}&=&-\frac{q r}{\varrho^2 \Xi}(d t-a sin^2\theta d
 \varphi)-\frac{p \cos \theta}{\varrho^2 \Xi}[a d t-(r^2+a^2) d
 \varphi], \label{aa}
\ea and the associated field strength tensor is \ba
 {\mathbf{F}}=&-&\frac{1}{\varrho^4}[q (r^2-a^2\cos^2\theta)+2
 p r a \cos \theta ]e^0\wedge e^1\nonumber\\
 &+&\frac{1}{\varrho^4}[q (r^2-a^2\cos^2\theta)-2 p r a \cos \theta
 ]e^2\wedge e^3.
\ea
 The Killing vector is
\be
 \chi_H^a = \partial_t + \Omega_H\partial_\varphi,
\ee
 where $\Omega_H=-\left(\frac{g_{t\varphi}}{g_{\varphi \varphi
}}\right)_H=\frac{\Xi a }{r_+^2+a^2}$ is the angular velocity of
the black hole. The analysis of the preceding subsection goes
through with virtually no changes, yields a statistical entropy
\begin{equation}\label{bs}
S=\frac{A_H}{4 G}=\frac{\pi}{G}\frac{r_+^2+a^2}{\Xi},
\end{equation}
which also coincides with its Bekenstein-Hawking entropy.

\vspace*{0.5cm}

\section{summary and discussion}

By using the covariant phase techniques we extend Carlip's
investigation in Ref. \cite{Carlip99}  to a case containing a
cosmological term and a electromagnetic field. If the event
horizon is treated as a boundary with Carlip's constraint
conditions\cite{Carlip99}, the central extension of the constraint
algebra is worked out, and a standard Virasoro subalgebra with a
corresponding central charge is constructed for the stationary
axisymmetric charged black hole. The statistical entropies of the
Kerr-Newman black hole and the Kerr-Newman-AdS black hole are then
obtained by using Cardy formula and the results agree with their
Bekenstein-Hawking entropies.

Since the static charged black holes, such as the
Reissner-Nordstr\"{o}m black hole and Reissner-Nordstr\"{o}m-AdS
black hole are special case of the metrics (\ref{metrick}) and
(\ref{metric}), respectively, the above results are also valid for
the static charged black holes.

The results obtained in this paper support Carlip's supposition:
regardless of the details of a quantum theory of gravity,
symmetries inherited from the classical theory may be sufficient
to determine the asymptotic behavior of the density of states.

\vspace*{2.5cm}

\begin{acknowledgements}
Jiliang Jing would like to thank Profs. Yongjiu Wang, Zhiming
Tang,  and Zongyang Sun for several helpful discussions. This work
was supported in part by the National Nature Science Foundation of
China under grant number 19975018.
\end{acknowledgements}

\newpage
\appendix

\section{Boundary Conditions}

In this section, we list the Carlip's boundary conditions
\cite{Carlip99} for convenience.

As Carlip did in Ref. \cite{Carlip99} we define a ``stretched
horizon"
\begin{equation}\label{chi}
  \chi ^2=\epsilon.
\end{equation}
where $\chi ^2=g_{ab}\chi^a\chi^b$, $\chi^a$ is a Killing vector.
The result of the computation will be evaluated at the event
horizon of the black hole by taking $\epsilon$ to zero. Near the
stretched horizon, one can introduce a vector orthogonal to the
orbit of $\chi^a$ by
\begin{equation}\label{def}
\nabla_a\chi^2=-2\kappa\rho_a,
\end{equation}
where $\kappa$ is the surface gravity. Vector $\rho^a$ satisfies
conditions
\begin{eqnarray}
 \chi^a\rho_a=-\frac{1}{\kappa}\chi^a\chi^b\nabla_a\chi_b=0,
 \qquad && \hbox{everywhere}\nonumber \\
 \rho^a\rightarrow \chi^a,\qquad && \hbox{at the horizon}
\end{eqnarray}

To preserve the ``asymptotic" structure at horizon, we impose
Carlip's boundary conditions\cite{Carlip99}
\begin{equation}\label{cond}
\delta \chi ^2=0,  \ \ \ \ \chi ^a t^b \delta g_{ab}=0, \ \ \ \
\delta \rho _a=-\frac{1}{2\kappa}\nabla _a(\delta \chi ^2)=0, \ \
\ \  at\ \ \ \chi^2=0,
\end{equation}
where $t^a$ is a any unit spacelike vector tangent to boundary
${\partial \mathbf{M}}$ of the spacetime ${\mathbf{M}}$.  And the
surface deformation vector is suggested as the following form
\begin{equation}\label{xi}
\xi^a ={\mathcal{R}}\rho^a+{\mathcal{T}}\chi ^a,
\end{equation}
where functions ${\mathcal{R}}$ and ${\mathcal{T}}$
satisfy\cite{Carlip99}
\begin{eqnarray}\label{rt}
{\mathcal{R}}=\frac{1}{\kappa}\frac{\chi^2}{\rho^2}\chi^a\nabla_a
{\mathcal{T}},\qquad && \hbox{everywhere} \nonumber \\
\rho^a\nabla_a{\mathcal{T}}=0, \qquad && \hbox{at the horizon}.
\end{eqnarray}
Fixing the average value of $\tilde{\kappa}$
($\tilde{\kappa}=-\frac{a^2}{\chi^2}$, $a^a=\chi^b\nabla_b\chi^a$
is the acceleration of an orbit of $\chi ^a$ ) over a cross
section of the horizon\cite{Carlip99}
\begin{equation}\label{value}
\delta\int_{\partial C}
\hat{\epsilon}\left(\tilde{\kappa}-\frac{\rho}{|\chi|}\kappa\right)=0,
\end{equation}
where $\kappa$ is the surface gravity and $\hat{\epsilon}$ is the
induced volume measure on ${\mathcal{H}}$ (${\mathcal{H}}$ denote
the (n-2)-dimensional intersection of the Cauchy surface $C$ with
the Killing horizon $\chi^2=0$). The technical role of the
condition (\ref{value}) is to guarantee the existence of
generators $H[\xi]$. For a one-parameter group of diffeomorphism
such that $D{\mathcal{T}}_\alpha=\lambda _\alpha{\mathcal{T}}_
\alpha$, ( $D\equiv \chi^a\partial_a$ ),  condition (\ref{value})
in turn implies an orthogonality relation\cite{Carlip99}
\begin{equation}\label{cond2}
\int_{\partial C}\hat{\epsilon}\
{\mathcal{T}}_\alpha{\mathcal{T}}_\beta \sim
\delta_{\alpha+\beta}.
\end{equation}
By using the other future-directed null normal vector
$
N^a=k^a-\alpha \chi^a-t^a, $  with
$k^a=-\frac{1}{\chi^2}\left(\chi^a-\frac{|\chi|}{\rho}
\rho^a\right)$ and a normalization $N_a\chi^a=-1$, the volume
element can be expressed as
\begin{equation}\label{vol}
\epsilon_{b c a_1 \cdots a_{n-2}}=\hat{\epsilon}_{ a_1 \cdots
a_{n-2}}(\chi_b N_c-\chi_c N_b)+\cdots\cdots,
\end{equation}
the omitted terms do not contribute to the integral.

\end{document}